\documentclass[10pt,superscriptaddress,aps,pre,twocolumn,nofootinbib]{revtex4-2}
\usepackage{amsmath,amssymb,amsthm}
\usepackage{multirow}
\usepackage{tabularx}
\usepackage[breaklinks=true,colorlinks,citecolor=blue,linkcolor=red,urlcolor=blue]{hyperref}
\usepackage[pdftex]{graphicx}
\usepackage{times,txfonts}
\usepackage{orcidlink}
\usepackage{braket}
\usepackage{color}
\usepackage{natbib}
\usepackage{url}
\usepackage{xurl}
\usepackage{makecell}
\usepackage{bm}
\usepackage{amsmath,blkarray}
\usepackage{mathtools}
\usepackage{makecell}
\usepackage[normalem]{ulem}
\usepackage{latexsym}
\usepackage{tabularx, booktabs}
\usepackage{graphics,epstopdf}
\usepackage{graphicx}
\usepackage[linesnumbered,lined,ruled,vlined]{algorithm2e}
\usepackage{amsfonts}
\usepackage{tikz}
\usepackage{graphicx}
\usepackage{subfigure} 

\usepackage{graphicx}
\usepackage{dcolumn}
\usepackage{bm}

\begin{document}

\title{Classical counterparts of shortcuts to adiabaticity in nonlinear dissipative Lagrangian systems\\}
\author{Jincheng Shi$^{\orcidlink{0009-0007-1297-6050}}$}
\affiliation{Department of Physics, Shanghai University, Shanghai 200444, China}

\author{Yicheng Pan$^{\orcidlink{0009-0001-9440-1600}}$}
\affiliation{Department of Engineering, King's College London, Strand, London WC2R 2LS, United Kingdom}

\author{Yue Ban$^{\orcidlink{0000-0003-1764-4470}}$}
\email{yue.ban@csic.es}
\affiliation{Instituto de Ciencia de Materiales de Madrid ICMM-CSIC, 28049 Madrid, Spain}

\author{Xi Chen$^{\orcidlink{0000-0003-4221-4288}}$}
\email{xi.chen@csic.es}
\affiliation{Instituto de Ciencia de Materiales de Madrid ICMM-CSIC, 28049 Madrid, Spain}

\date{\today}

\begin{abstract}
Shortcuts to adiabaticity (STA) were first developed in quantum dynamics to realize rapid transformations with suppressed residual excitations. Here we show how the same idea can be implemented in classical nonlinear dissipative Lagrangian systems. Using a coupled $r$-$\theta$ manipulator as an illustrative model,  we perform inverse engineering on the Euler-Lagrange equations with Rayleigh dissipation by prescribing endpoint-stationary trajectories, obtaining the corresponding force and torque profiles and quantifying how geometric coupling amplifies errors and residual energy. We further compare smooth STA protocols with actuator-bounded time-optimal solutions and with proportional–integral–derivative tracking, which highlights a trade-off among smoothness, speed, and robustness. Finally, we introduce a single-shot correction based on one mid-course measurement to reduce the effect of early deviations while keeping the inputs nearly smooth. These results provide a practical bridge between quantum STA concepts and their classical counterparts.
\end{abstract}

\maketitle


\section{Introduction}

In recent years, a variety of approaches under the umbrella of ``shortcuts to adiabaticity'' (STA) have been developed to accelerate slow adiabatic processes while suppressing residual excitations in atomic, molecular, and optical physics \cite{review1,review2}. Representative strategies include fast-forward scaling \cite{masuda1,masuda2,Torrontegui1}, counter-diabatic driving (transitionless quantum driving) \cite{Rice,Rice2,berry,chenprl105,Deffner1,del1}, and invariant-based inverse engineering \cite{chensta}. Invariant-based inverse engineering, in particular, has been widely explored both theoretically and experimentally for fast frictionless cooling \cite{chensta,nice1,nice2} and for transport in time-dependent harmonic traps \cite{DavidEPL,Torrontegui2,Palmero2,KimNC,Corgier,Hickman}. Because only boundary conditions are fixed while the intermediate evolution retains design freedom, STA protocols can be further optimized using Pontryagin's minimum principle (PMP) with respect to time, residual excitation energy, or other cost functions \cite{Lipra10,Lupra14,Chenpra11,Lupra142,Ding}. Beyond quantum platforms, inverse-engineering ideas have also been extended to mechanical resonators \cite{Liyong}, photonic lattices \cite{Stefanatos1}, polarization retarders \cite{WangOL23}, classical RC circuits \cite{Faure1,RC25}, and even biological contexts \cite{NP2020}. More generally, engineered fast protocols can be formulated for systems governed by dynamical equations without simple invariants, such as Langevin and Fokker--Planck descriptions in nonequilibrium statistical mechanics \cite{Mart1}.

A natural perspective behind many STA constructions is that one first prescribes boundary-conditioned trajectories and then reconstructs the drives required to realize them. {In standard STA formulations, this idea is often implemented through dynamical invariants or counter-diabatic driving, which provide a clear reference for the quantity that remains effectively adiabatic during the evolution \cite{ChenMugapra11}. For systems admitting Lagrangian descriptions \cite{Cirac,Cirac2}, however, the same inverse viewpoint can be formulated directly at the level of the Euler-Lagrange (EL) equations, which relate generalized coordinates and generalized forces in a structured way. In conservative or weakly nonlinear settings, this Lagrangian formulation can be closely connected to the conventional invariant-based picture \cite{li2016shortcut,Chaos,Tianniu,LiOE}. By contrast, in nonlinear dissipative systems explicit dynamical invariants are often unavailable, and the EL equations themselves provide a natural auxiliary description from which excitation-suppressed finite-time trajectories can be constructed, either exactly or approximately. This suggests a broader interpretation of STA in classical mechanics: rather than relying exclusively on explicit invariants, one may use the equations of motion to engineer finite-time protocols that reproduce the excitation-suppressed outcome that would result from slow adiabatic evolution.}

Quantum-classical analogies in this direction are conceptually useful as well: classical adiabatic invariants and adiabatic approximations have long been known, and they naturally motivate finite-time protocols that suppress unwanted excitations. Along these lines, Jarzynski and coworkers proposed classical versions of counter-diabatic driving \cite{jarzynski1} and fast-forward protocols for classical adiabatic invariants \cite{jarzynski2}. Okuyama and Takahashi formulated STA in classical mechanics and constructed counter-diabatic terms from the dispersionless Korteweg--de Vries hierarchy \cite{Kazu}. Moverover, invariant-based ideas and their variants have also been applied to classical models such as cranes \cite{Gonz1,cranerobust}, gun-turret dynamics \cite{stefanatos}, and cartpole systems \cite{Kongentropy,Ion}, while classical Lagrangian mechanics can in turn inform STA design when strong nonlinearities, such as anharmonicity, are present \cite{Lianharmonic}.  {Such a viewpoint is also practically appealing. Once a reference trajectory is constructed from the EL dynamics, actuator constraints can be incorporated, comparisons with time-optimal strategies can be performed, and feedback or correction steps can be introduced when needed. This practical route is particularly attractive for nonlinear dissipative Lagrangian systems, where invariant-based approaches are difficult to implement.}

\begin{figure}[t]
	\centering
	\includegraphics[width=2.2in]{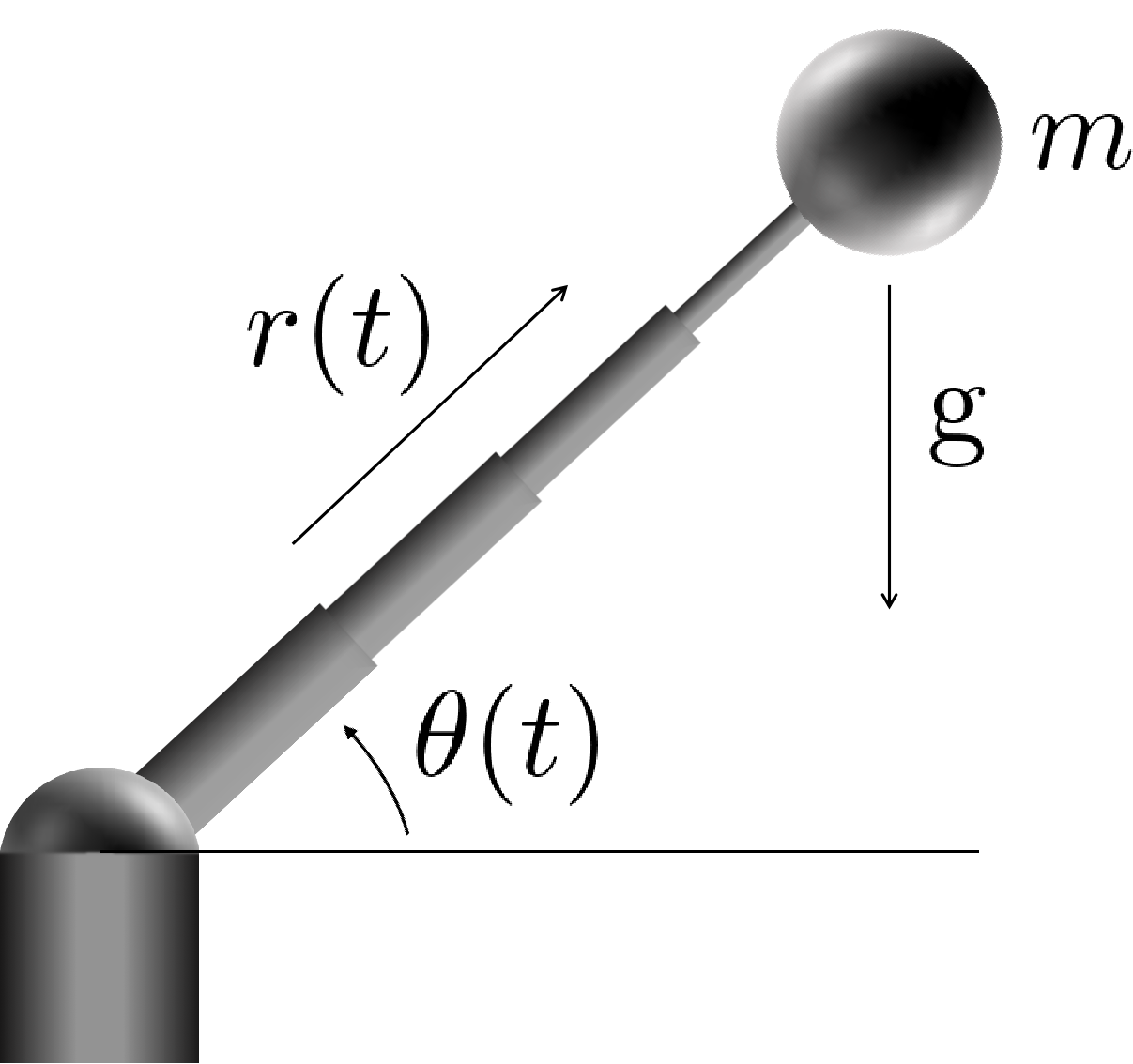}
	\caption{Schematic diagram of an $r$-$\theta$ manipulator used as a simple illustrative model,  consisting of a revolute joint $\theta(t)$, a prismatic arm with time-dependent length $r(t)$, and a point mass $m$, in the presence of gravity $g$.}
	\label{fig:model}
\end{figure}

In this work, we develop an inverse-engineering STA framework for classical nonlinear dissipative Lagrangian systems and use a coupled $r$-$\theta$ manipulator (Fig.~\ref{fig:model}) as an illustrative model capturing the geometric coupling inherent to polar-coordinate dynamics. We first derive the coupled equations of motion from the EL equations with Rayleigh dissipation. We then prescribe endpoint-stationary trajectories for the generalized coordinates and reconstruct the corresponding generalized force and torque profiles that implement fast finite-time transfers while suppressing residual excitations at the final time. We analyze robustness against parameter mismatch and disturbances, and benchmark the smooth STA protocols against actuator-bounded time-optimal solutions and PID tracking, thereby revealing the trade-offs among smoothness, speed, and robustness. Finally, we introduce a single-shot correction based on a mid-course measurement to mitigate the propagation of early deviations while preserving nearly smooth control inputs.

\section{LAGRANGIAN MECHANICS for $r$-$\theta$ manipulator }
\label{model}

In Fig.~\ref{fig:model}, we introduce a minimal $r$-$\theta$ model as an illustrative nonlinear dissipative Lagrangian system with geometric coupling in polar coordinates. {It consists of a massless arm with two degrees of freedom: the revolute angle $\theta(t)$ and the radial length $r(t)$, with a point mass $m$ moving under gravity $g$ and subject to viscous losses.} Our goal is to realize a rapid transfer from an initial configuration [$r(0)\equiv r_0,\theta(0) \equiv \theta_0]$ to a target configuration $[r(t_f)\equiv r_f,\theta(t_f)\equiv  \theta_f]$ within a prescribed duration $t_f$, while suppressing residual energies at the initial and final times.

The generalized coordinates are $q(t)=[\theta(t),r(t)]^{T}$. The corresponding generalized inputs are the torque $\tau(t)$ and radial force $f(t)$, i.e., $Q=[\tau,f]^T$. The Lagrangian reads
\begin{equation}
L=T-V=\tfrac{1}{2}mr^2\dot{\theta}^2+\tfrac{1}{2}m\dot{r}^2-mgr\sin\theta,
\end{equation}
where the first two terms are the rotational and radial kinetic energies and the last term is the gravitational potential. Viscous dissipation is modeled by a Rayleigh dissipation function,
\begin{equation}
P=\tfrac{1}{2}B_1\dot{\theta}^2+\tfrac{1}{2}B_2\dot{r}^2,
\label{eqdissipation}
\end{equation}
with damping coefficients $B_1$ and $B_2$ for the angular and radial motions, respectively. The equations of motion follow from the EL equations for a nonconservative system with Rayleigh dissipation \cite{galleycr},
\begin{equation}
\frac{d}{dt}\!\left(\frac{\partial L}{\partial \dot q_i}\right)-\frac{\partial L}{\partial q_i}+\frac{\partial P}{\partial \dot q_i}=Q_i,
\end{equation}
which yield the coupled nonlinear dynamics
\begin{align}
mr^2\ddot{\theta}+2mr\dot{r}\dot{\theta}+B_1\dot{\theta}+mgr\cos\theta&=\tau, \label{eq:theta_dyn}\\
m\ddot{r}-mr\dot{\theta}^2+B_2\dot{r}+mg\sin\theta&=f. \label{eq:r_dyn}
\end{align}
Here and in what follows, explicit time dependence is omitted for brevity.

In compact robotic form, Eqs.~(\ref{eq:theta_dyn}) and (\ref{eq:r_dyn}) can be written as
\begin{equation}
\label{Meq}
M(q)\ddot{q}+C(q,\dot{q})\dot{q}+G(q)+D\dot{q}=Q,
\end{equation}
with the inertia and damping matrices
\[
M(q)=
\begin{bmatrix}
mr^2 & 0\\
0 & m
\end{bmatrix},\qquad
D=
\begin{bmatrix}
B_1 & 0\\
0 & B_2
\end{bmatrix},
\]
and the Coriolis/centripetal and gravity contributions
\[
C(q,\dot{q})\dot{q}=
\begin{bmatrix}
2mr\dot r\dot\theta\\
-mr\dot\theta^2
\end{bmatrix},\qquad
G(q)=
\begin{bmatrix}
mgr\cos\theta\\
mg\sin\theta
\end{bmatrix}.
\]

{The target configuration is denoted by 
$ q(t_f)\equiv q_f=(\theta_f,r_f) $,
which corresponds to the desired final state of the transfer. 
This configuration becomes a stationary equilibrium of the dynamics when the generalized inputs balance gravity, i.e., $Q_f=G(q_f)$. 
To analyze the local behavior near this equilibrium, we introduce small deviations
$\theta=\theta_f+\delta\theta$ and $r=r_f+\delta r$.
Linearizing Eq.~(\ref{Meq}) around $q_f$ yields a second-order system of the form
\begin{equation}
M\,\ddot{\delta q}+D\,\dot{\delta q}+K_{\mathrm{eff}}\,\delta q=0,
\end{equation}
where $\delta q=(\delta\theta,\delta r)^T$ and 
$K_{\mathrm{eff}}=\partial G/\partial q|_{q_f}$ 
is the effective stiffness matrix obtained from the linearization of the gravitational and geometric terms. The eigenmodes of this linearized system describe small oscillations of the manipulator around the target configuration.
In the absence of dissipation ($D=0$), this linearized dynamics reduces to a system of coupled harmonic oscillators. In that limit the normal modes admit classical action invariants, which are approximately conserved under slow parameter variations.
The endpoint conditions imposed in the inverse-engineering protocol,
\begin{equation}
\label{boundaryq}
\dot q(0)=\dot q(t_f)=0, \qquad 
\ddot q(0)=\ddot q(t_f)=0,
\end{equation}
ensure that the system reaches the final configuration $q(t_f)=q_f$ with vanishing velocity and acceleration. In the linearized description this corresponds to suppressing the excitation amplitudes of the oscillatory modes, so that no residual motion remains near the equilibrium point. Although this construction does not rely on an explicit adiabatic invariant as in standard STA formulations \cite{jarzynski1,jarzynski2}, it reproduces the physical outcome of a slow adiabatic evolution: the system arrives at the equilibrium configuration without exciting dynamical modes. In this sense, the boundary-conditioned inverse-engineering protocol provides a general classical analogue of STA in the operational sense, where a finite-time protocol that reproduces the excitation-suppressed outcome without requiring slow driving.}

Although we use the $r$-$\theta$ manipulator as an illustrative example, the inverse-engineering procedure applies to generic multi-DOF robotic systems within the same Lagrangian structure (\ref{Meq}), including standard planar two-link (2R) manipulators. In the following, we construct STA protocols via inverse engineering and benchmark them against actuator-bounded time-optimal solutions and feedback tracking, thereby demonstrating fast transfers with suppressed terminal excitations and clarifying the trade-offs among smoothness, speed, and robustness in coupled nonlinear dissipative dynamics.

\section{STA and Optimization}
\label{STA&Optimization}

\subsection{Inverse engineering}
\label{STA}

In this section, we construct STA protocols for the coupled nonlinear Lagrangian dynamics by using inverse engineering of the time-dependent generalized force and torque. The basic idea is to prescribe a smooth finite-time evolution of the generalized coordinates and then reconstruct the driving terms from the equations of motion. To this end, we impose endpoint-stationary boundary conditions on both coordinates from Eqs. (\ref{boundaryq}). For the angular coordinate,
\begin{eqnarray}
\begin{array}{l}
\theta(0)=\theta_0,\quad \theta(t_f)=\theta_f,\\
\dot\theta(0)=\ddot\theta(0)=\dot\theta(t_f)=\ddot\theta(t_f)=0,
\end{array}
\label{eq6}
\end{eqnarray}
and for the radial coordinate,
\begin{eqnarray}
\begin{array}{l}
r(0)=r_0,\quad r(t_f)=r_f,\\
\dot r(0)=\ddot r(0)=\dot r(t_f)=\ddot r(t_f)=0.
\end{array}
\label{eq7}
\end{eqnarray}
These conditions enforce endpoint stationarity and eliminate residual oscillations at the beginning and end of the protocol.

To satisfy Eqs.~(\ref{eq6}) and (\ref{eq7}), we parameterize $q(t)=[\theta(t),r(t)]^{T}$ using a quintic polynomial ansatz,
\begin{eqnarray}
q_i(t)=q_{0i}+d_i\left(6s^5-15s^4+10s^3\right),
\label{eqquintic}
\end{eqnarray}
where $s=t/t_f$ is the normalized time, $q_0=[\theta_0,r_0]^T$ and $q_f=[\theta_f,r_f]^T$ are the boundary configurations, $d=q_f-q_0$, and $i\in\{\theta,r\}$. This choice automatically fulfills the position, velocity, and acceleration constraints and yields a smooth trajectory. Differentiation gives
\begin{eqnarray}
\dot q_i(t)=\frac{30d_i}{t_f}\left(s^4-2s^3+s^2\right),
\label{eqdotquintic}
\end{eqnarray}
and
\begin{eqnarray}
\ddot q_i(t)=\frac{60d_i}{t_f^2}\left(2s^3-3s^2+s\right).
\label{eqddotquintic}
\end{eqnarray}
Substituting Eqs.~(\ref{eqquintic})--(\ref{eqddotquintic}) into the equations of motion (\ref{eq:theta_dyn}) and (\ref{eq:r_dyn}) yields the required time-dependent torque $\tau(t)$ and radial force $f(t)$ that implement the prescribed STA trajectory. Since the quintic ansatz is smooth up to the second derivative, the resulting $\tau(t)$ and $f(t)$ are continuous in time, avoiding discontinuous actuation.

\begin{figure}[t]
\centering
\includegraphics[width=\linewidth]{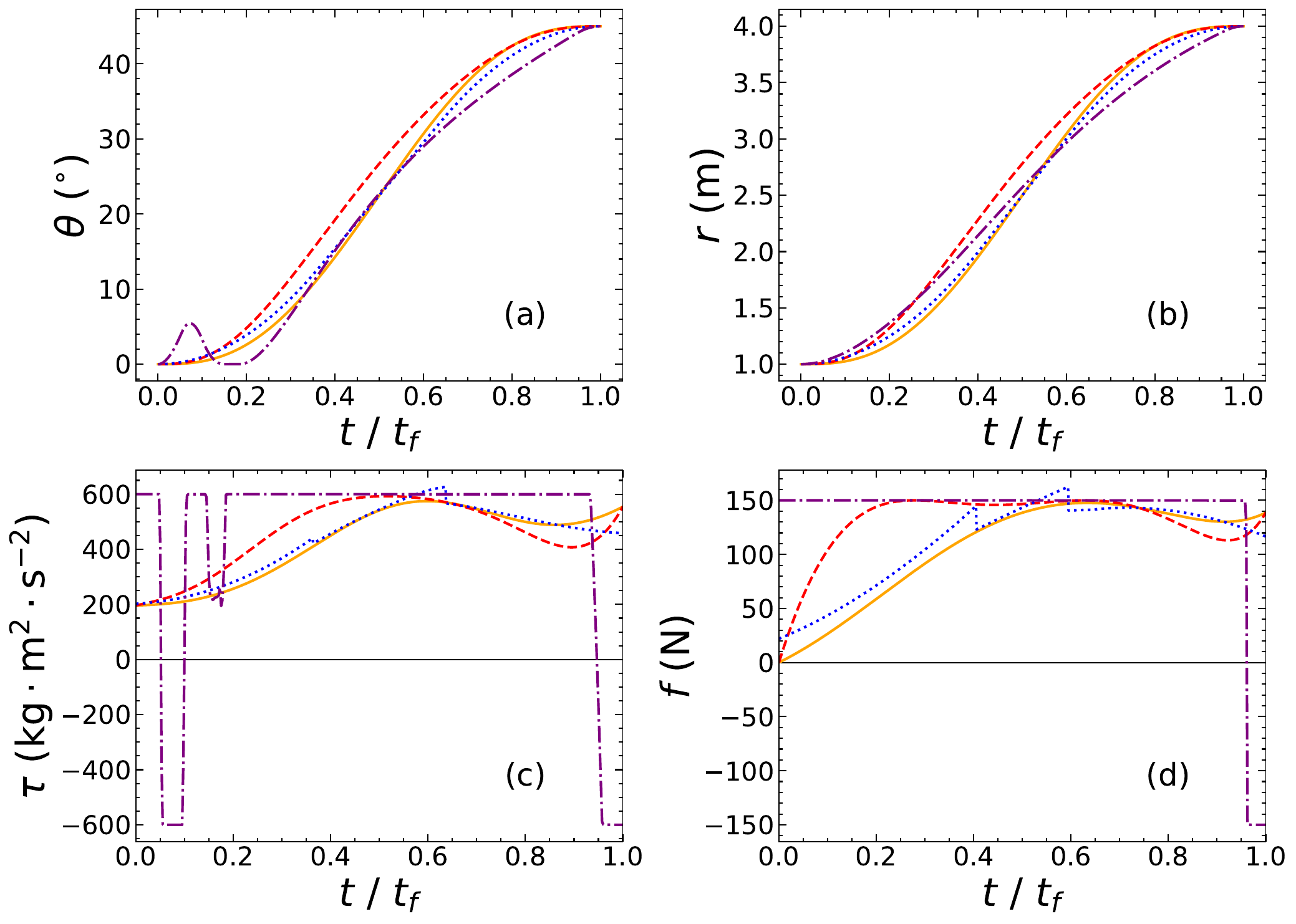}
\caption{\label{figstaopt}(a) Angular coordinate $\theta(t)$, (b) radial coordinate $r(t)$, (c) generalized torque $\tau(t)$, and (d) generalized force $f(t)$ for different protocols.
STA trajectories are generated using the quintic polynomial (\ref{eqquintic}) (solid orange, $t_f=4~\mathrm{s}$) and the seventh-order polynomial (\ref{eqseventh}) (dashed red, $t_f=2.535~\mathrm{s}$).
They are compared with a constraint-limited protocol satisfying the kinematic bounds (\ref{eq15}) (dotted blue, $t_f=3.364~\mathrm{s}$) and with the actuator-bounded time-optimal protocol under (\ref{eq17}) (dash-dotted purple, $t_f=1.755~\mathrm{s}$).
The transfer is from $(\theta_0,r_0)=(0^\circ,1~\mathrm{m})$ to $(\theta_f,r_f)=(45^\circ,4~\mathrm{m})$ with the parameters such as $m=20~\mathrm{kg}$, $g=9.8~\mathrm{m\,s^{-2}}$, $B_1=100~\mathrm{kg\,m^2\,s^{-1}\,rad^{-1}}$, and $B_2=50~\mathrm{kg\,s^{-1}}$.
}
\end{figure}

To illustrate inverse engineering, we consider a representative transfer between two configurations: $\theta_0=0^\circ$, $\theta_f=45^\circ=\pi/4~\mathrm{rad}$, and $r_0=1~\mathrm{m}$, $r_f=4~\mathrm{m}$, with $t_f=4~\mathrm{s}$. The parameters are $m=20~\mathrm{kg}$, $g=9.8~\mathrm{m\,s^{-2}}$, and damping coefficients $B_1=100~\mathrm{kg\,m^2\,s^{-1}\,rad^{-1}}$ and $B_2=50~\mathrm{kg\,s^{-1}}$. Using the trajectories in Eqs.~(\ref{eqquintic})--(\ref{eqddotquintic}), the corresponding $\tau(t)$ and $f(t)$ follow from inverse substitution into the dynamics. The resulting motions and generalized forces are shown by solid orange curves in Fig.~\ref{figstaopt}.

In principle, the protocol duration $t_f$ can be chosen freely. In practice, however, feasible motions are restricted by kinematic limits, such as bounds on velocities and accelerations, which impose a lower bound on admissible transfer times. To examine this constraint-limited regime, we impose
\begin{eqnarray}
\label{eq15}
\begin{aligned}
& |\dot{\theta}(t)| \leq 0.4~\mathrm{rad\,s^{-1}}, 
&& |\ddot{\theta}(t)| \leq 0.3~\mathrm{rad\,s^{-2}}, \\
& |\dot{r}(t)| \leq 1.5~\mathrm{m\,s^{-1}}, 
&& |\ddot{r}(t)| \leq 1.2~\mathrm{m\,s^{-2}}.
\end{aligned}
\end{eqnarray}
The STA trajectories satisfy these bounds for sufficiently large $t_f$. As $t_f$ decreases, the trajectory approaches the constraint boundaries; below a critical duration, Eq.~(\ref{eq15}) cannot be satisfied within the chosen trajectory class. {The bounds in Eqs.~(\ref{eq15}) and (\ref{eq17}) represent typical kinematic and actuator limits in robotic manipulators \cite{siciliano2009}, while the specific numerical values are chosen for illustrative purposes.}

To construct the fastest feasible protocol within these kinematic limits, we adopt a simple inverse-design strategy based on piecewise acceleration profiles {of the bang-off-bang form \cite{stefanatos}} consistent with Eq.~(\ref{eq15}).  The motion first accelerates at the maximal admissible acceleration until the velocity bound is reached, then evolves with a coasting phase, and finally decelerates to satisfy the endpoint conditions. The corresponding $\tau(t)$ and $f(t)$ are obtained by inverse substitution into the equations of motion.  {Because the two coordinates may reach their targets at different times under these bounds, we synchronize their arrival by adjusting the coasting phase of the faster coordinate. In practice, the coasting velocity of the leading coordinate is reduced while keeping its acceleration and deceleration phases within the imposed limits. This modification preserves the bang–off–bang structure and ensures that both velocity and acceleration constraints are satisfied, while the final arrival time is determined by the slower coordinate.} With this synchronization, the shortest feasible duration within the adopted profile class is $t_f=3.364~\mathrm{s}$ (set by the radial motion), as shown by the dotted blue curves in Fig.~\ref{figstaopt}. Compared with the smooth STA protocols, the constraint-limited construction typically produces piecewise accelerations and hence nonsmooth generalized forces, highlighting a trade-off between strict kinematic feasibility and smooth actuation.

\subsection{Time-optimal protocol}

In many situations, the most relevant constraints act directly on the actuator outputs rather than on kinematic variables. We therefore impose bounds on the generalized torque and force,
\begin{eqnarray}
|\tau(t)|\le 600~\mathrm{kg\,m^2\,s^{-2}},\qquad |f(t)|\le 150~\mathrm{N},
\label{eq17}
\end{eqnarray}
and seek the minimum-time transfer consistent with these limits. PMP provides a standard framework for this time-optimal problem, with the cost functional
\begin{eqnarray}
J=\int_{0}^{t_f}1\,dt=t_f.
\end{eqnarray}
We define the state vector $\mathbf{x}=[\theta,r,\dot\theta,\dot r]^T$ and the control vector $\mathbf{u}=[\tau,f]^T$, and write the dynamics compactly as $\dot{\mathbf{x}}=\mathbf{f}(\mathbf{x},\mathbf{u})$. The control Hamiltonian is
\begin{eqnarray}
H_c(\mathbf{p},\mathbf{x},\mathbf{u})=1+\mathbf{p}^T\mathbf{f}(\mathbf{x},\mathbf{u}),
\end{eqnarray}
where $\mathbf{p}(t)=[p_1,p_2,p_3,p_4]^T$ is the costate. The state--costate equations are
$\dot{\mathbf{x}}=\partial H_c/\partial \mathbf{p}$ and
$\dot{\mathbf{p}}=-\partial H_c/\partial \mathbf{x}$.
PMP states that the optimal control $\mathbf{u}^*(t)=[\tau^*(t),f^*(t)]^T$ minimizes $H_c$ pointwise,
\begin{eqnarray}
\mathbf{u}^*(t)=\arg\min_{\mathbf{u}\in U}H_c(\mathbf{p}(t),\mathbf{x}(t),\mathbf{u}),
\label{eqpmp}
\end{eqnarray}
where $U$ is the admissible set defined by Eq.~(\ref{eq17}).

Since the dynamics are affine in $\tau$ and $f$, the Hamiltonian can be written as
\begin{eqnarray}
H_c(\mathbf{p},\mathbf{x},\mathbf{u})=\Phi(\mathbf{p},\mathbf{x})+\alpha_\tau(\mathbf{p},\mathbf{x})\,\tau+\alpha_f(\mathbf{p},\mathbf{x})\,f,
\end{eqnarray}
where $\Phi$ collects the terms independent of the controls {and the coefficients $\alpha_{\tau}=p_3/mr^2$ and $\alpha_{f}=p_4/m$ play the role of switching functions in the Pontryagin framework.} Consequently, the minimizer typically lies on the bounds, leading to a bang-bang structure,
\begin{eqnarray}
\tau^*(t) &=& 
\begin{cases}
\tau_{\text{max}}, & p_3/m r^2 < 0 \\
-\tau_{\text{max}}, & p_3/m r^2  > 0
\end{cases}, \\
f^*(t) &=& 
\begin{cases}
f_{\text{max}}, & p_4/m < 0 \\
-f_{\text{max}}, & p_4/m > 0
\end{cases}.
\end{eqnarray}
In practice, however, the torque channel can deviate from an ideal bang–bang waveform because $\tau$ enters the coupled state–costate dynamics. {In particular, the costate equation for the radial coordinate contains terms involving $\tau$, which influence the evolution of the switching function $\alpha_\tau$. Together with the nonlinear geometric coupling, gravity, and dissipation, this interaction may lead to intervals where the optimal torque deviates from pure bang–bang switching and exhibits smoother modulation.} A closed-form solution is therefore not available, particularly because the final time is not known \emph{a priori}. We solve the resulting boundary-value problem numerically in \textsc{Matlab} using \textsc{CasADi}. The obtained protocol satisfies the boundary conditions and the actuator bounds in Eq.~(\ref{eq17}) and is shown by the purple dash-dotted curves in Fig.~\ref{figstaopt}. Consistent with the above discussion, the optimal force $f(t)$ remains close to bang-bang, whereas the optimal torque $\tau(t)$ displays noticeable deviations due to gravity and damping.

The time-optimal trajectory for $\theta(t)$ also differs markedly from the smooth STA trajectories and may exhibit pronounced initial oscillations. Numerical tests with different discretizations and initial guesses indicate that allowing stronger oscillations can further reduce the optimal duration. {Physically, the initial oscillations in $\tau(t)$ can be interpreted as a mechanism to rapidly modulate the angular velocity. By briefly increasing and then reducing $\dot{\theta}$, the system generates a strong centrifugal term $-mr\dot{\theta}^2$ in the radial equation, which assists the radial actuator in extending the arm more quickly. In this way, the control exploits the geometric coupling between the angular and radial motions to accelerate the radial expansion and reduce the overall transfer time.} In what follows, we report a representative solution that balances smoothness and speed and achieves $t_f=1.755~\mathrm{s}$. The endpoint acceleration constraints are enforced through rapid admissible switching, while the position and velocity boundary conditions are satisfied exactly.

\begin{figure}[t]
\centering
\includegraphics[width=\linewidth,height=3cm]{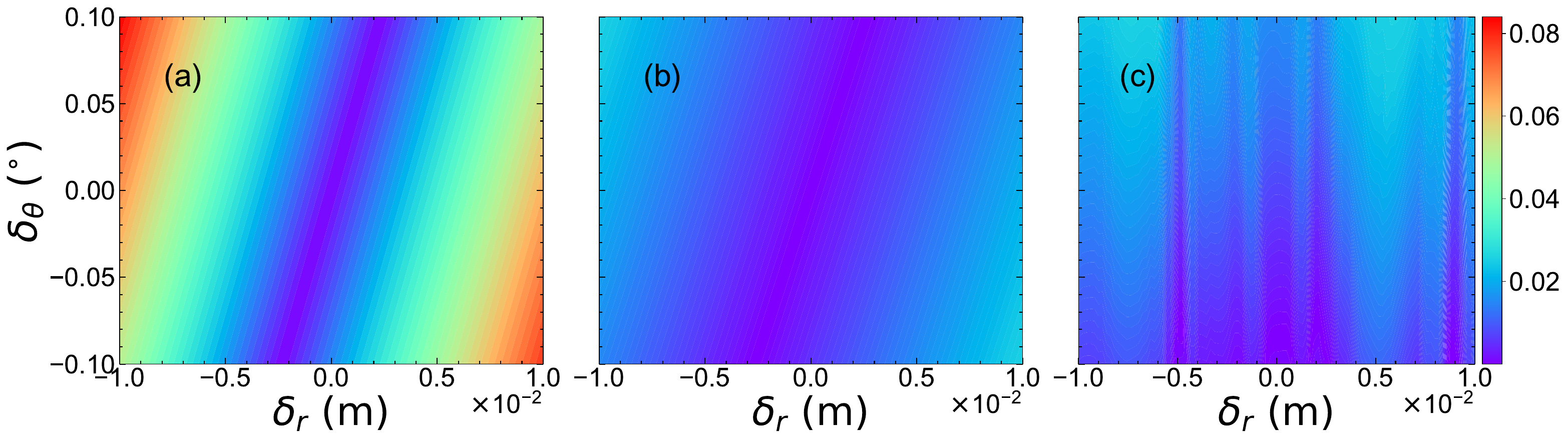}
\caption{\label{figinitialerror}Relative error $\mathrm{RE}$ of the final mechanical energy as a function of initial-state errors $(\delta_\theta(0),\delta_r(0))$.
(a) Smooth STA protocol based on the seventh-order trajectory (\ref{eqseventh}) with $t_f=2.535~\mathrm{s}$.
(b) Actuator-bounded time-optimal protocol under (\ref{eq17}) with $t_f=1.755~\mathrm{s}$.
(c) PID tracking implemented with sampling time $\Delta t=0.01~\mathrm{s}$, using component-wise gains
$K_p^{(\theta)}=2.1\times10^{4}$, $K_i^{(\theta)}=1.5\times10^{4}$, $K_d^{(\theta)}=70$ for the torque channel and
$K_p^{(r)}=2.1\times10^{4}$, $K_i^{(r)}=1.5\times10^{4}$, $K_d^{(r)}=70$ for the force channel.
The torque-gain units are $\mathrm{kg\,m^2\,s^{-2}\,rad^{-1}}$ and the force-gain units are $\mathrm{kg\,s^{-2}}$.
Initial errors are sampled over $\delta_\theta(0)\in[-\pi/1800,\pi/1800]~\mathrm{rad}$ and $\delta_r(0)\in[-0.01,0.01]~\mathrm{m}$.
}
\end{figure}

\subsection{Comparison and discussion}

To compare smooth inverse-engineering protocols with actuator-bounded time-optimal solutions under identical limits, we impose the bounds on $\tau(t)$ and $f(t)$ in Eq.~(\ref{eq17}) and ask how short the STA transfer can be made while retaining a prescribed smooth trajectory class. Starting from the quintic trajectory (\ref{eqquintic}), the smallest feasible duration under Eq.~(\ref{eq17}) is $t_f=3.615~\mathrm{s}$, which is more than twice the duration of the PMP time-optimal solution. This gap reflects a basic trade-off: boundary-conditioned smooth trajectories restrict the admissible waveforms of the generalized forces, whereas time-optimal solutions exploit actuator saturation and rapid switching to minimize $t_f$.

To reduce $t_f$ within the STA framework, we enlarge the admissible trajectory space. Higher-order polynomials provide additional degrees of freedom \cite{entropyQi} that redistribute acceleration and jerk in time, thereby reshaping the required $\tau(t)$ and $f(t)$ and delaying saturation under the same actuator bounds. We therefore adopt the seventh-order ansatz
\begin{eqnarray}
q_i(t)=q_{0i}+d_i\left(20s^{7}-76s^{6}+114s^{5}-83s^{4}+26s^{3}\right),
\label{eqseventh}
\end{eqnarray}
which satisfies the same endpoint conditions while offering a more flexible acceleration profile. As shown in Fig.~\ref{figstaopt}, enforcing Eq.~(\ref{eq17}) reduces the minimum feasible STA duration to $t_f=2.535~\mathrm{s}$, substantially closer to the PMP optimum. Importantly, the corresponding $\tau(t)$ and $f(t)$ (dashed red) remain continuous, in contrast to the rapid switching typical of time-optimal solutions. In the remainder of this section, we take the seventh-order STA protocol as the representative smooth inverse-engineering baseline.

Beyond transfer time, robustness is central for coupled nonlinear dynamics because open-loop protocols cannot correct deviations once the evolution starts. Because endpoint stationarity targets vanishing residual kinetic energy, an energy-based terminal metric provides a compact proxy for residual motion beyond mere position error. Thus, to quantify terminal deviations, we define the mechanical energy 
\begin{equation}
E(t)=T(t)+V(t)=\tfrac12 mr^2\dot\theta^2+\tfrac12 m\dot r^2+mgr\sin\theta.
\end{equation}
For a given protocol, we define the nominal (error-free) final energy as $E_f \equiv E (t_f)$, {the ideal mechanical energy at the target final state,} obtained by propagating the dynamics with the same input $\{\tau(t),f(t)\}$ from the nominal initial condition $(\theta_0,r_0,\dot\theta_0,\dot r_0)$.
For perturbed initial conditions $(\theta_0+\delta_\theta(0),\, r_0+\delta_r(0),\,0,\,0)$, we denote the resulting final energy by
$ E_\delta \equiv E_{\rm pert}(t_f)$. The relative energy error reported in Fig.~\ref{figinitialerror} is then
\begin{equation}
\mathrm{RE}=\left|\frac{E_\delta-E_f}{E_f}\right|.
\end{equation}
Moreover,  Fig.~\ref{figinitialerror} shows the resulting $\mathrm{RE}$ maps over $\delta_\theta(0)\in[-\pi/1800,\pi/1800]~\mathrm{rad}$ and $\delta_r(0)\in[-0.01,0.01]~\mathrm{m}$. Both open-loop protocols exhibit similar qualitative structures, indicating comparable dominant sensitivity directions. Quantitatively, the time-optimal protocol yields a smaller mean relative error ($\mathrm{MRE}$) of $1.063\%$, compared with $3.372\%$ for the smooth STA protocol. For the present parameter set and bounds, this improvement is consistent with the shorter transfer time, which leaves less time for initial deviations to be converted into terminal errors by the nonlinear $r$-$\theta$ coupling. We stress that this advantage is scenario-dependent and reflects the combined effects of duration, waveform structure, and nonlinearity.

\begin{figure}[t]
\centering
\includegraphics[width=0.87\linewidth]{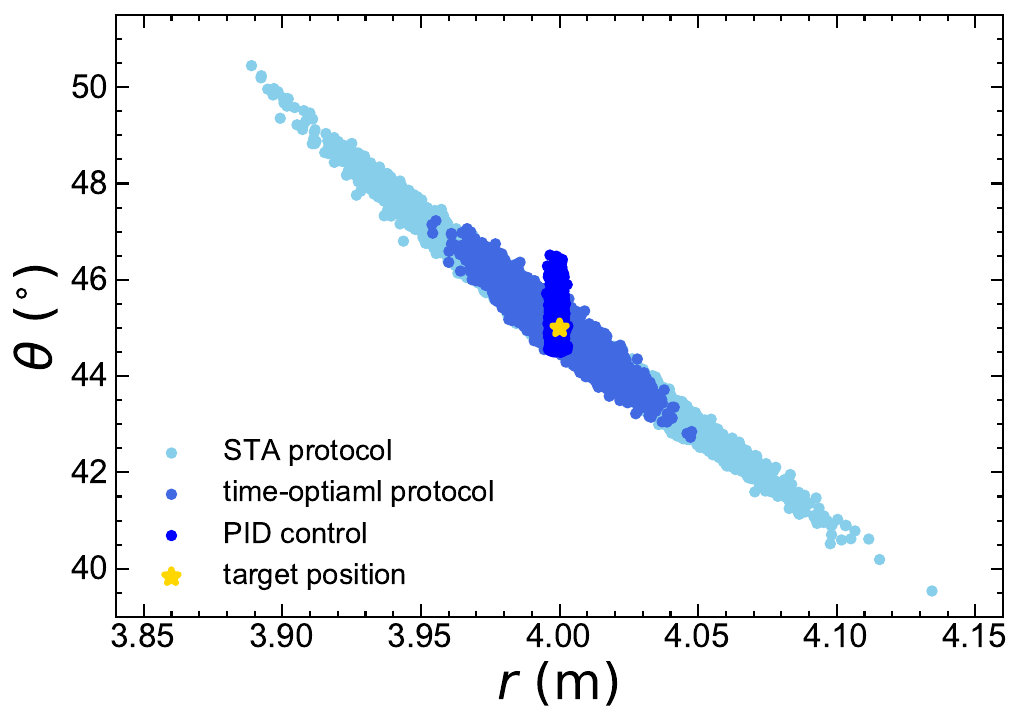}
\caption{\label{figcontrolerror}Scatter of the final configurations $(\theta(t_f),r(t_f))$ over repeated trials for the STA protocol, the actuator-bounded time-optimal protocol, and PID tracking. Zero-mean Gaussian white noise is added to the inputs, with $\delta_\tau(t)\in[-30,30]~\mathrm{kg\,m^2\,s^{-2}}$ and $\delta_f(t)\in[-10,10]~\mathrm{N}$. All other parameters are identical to those used in Fig.~\ref{figinitialerror}.
}
\end{figure}

To probe sensitivity to actuation fluctuations, we add zero-mean Gaussian white-noise perturbations $\delta_\tau(t)$ and $\delta_f(t)$ to the generalized torque and force,
\begin{align}
mr^2\ddot{\theta}+2mr\dot{r}\dot{\theta}+B_1\dot{\theta}+mgr\cos\theta&=\tau+\delta_\tau,\\
m\ddot{r}-mr\dot{\theta}^2+B_2\dot{r}+mg\sin\theta&=f+\delta_f.
\end{align}
We take $\delta_\tau\in[-30,30]~\mathrm{kg\,m^2\,s^{-2}}$ and $\delta_f\in[-10,10]~\mathrm{N}$ and perform repeated trials. The final-position scatter in Fig.~\ref{figcontrolerror} shows that both open-loop protocols are sensitive to input disturbances, as expected in the absence of feedback.

Quantitatively, the time-optimal protocol yields a slightly smaller $\mathrm{MRE}$ ($0.598\%$) than the STA protocol ($1.307\%$). A plausible interpretation is that the near-saturated inputs in the time-optimal solution reduce the relative impact of small additive perturbations, whereas smoother STA inputs operate closer to the actuator margins over longer times and can accumulate noise-induced deviations. Together, these results highlight a trade-off between smoothness and speed in coupled nonlinear dissipative dynamics under actuator constraints.

\section{Closed-loop control}

\subsubsection{PID tracking}

The results above highlight a generic limitation of open-loop protocols in coupled nonlinear dynamics: performance degrades in the presence of model mismatch and disturbances because deviations cannot be corrected once the evolution starts. As a benchmark for feedback stabilization, we introduce a proportional-integral-derivative (PID) tracking scheme \cite{pid}. We take the smooth seventh-order STA trajectory (\ref{eqseventh}) with $t_f=2.535~\mathrm{s}$ as a reference and formulate a trajectory-tracking problem for the dynamics (\ref{eq:theta_dyn}) and (\ref{eq:r_dyn}).

We apply PID feedback independently to the two actuation channels. Denoting the reference trajectory by $(\theta_{\rm ref}(t),r_{\rm ref}(t))$ and the tracking errors by
$e_\theta(t)=\theta_{\rm ref}(t)-\theta(t)$ and $e_r(t)=r_{\rm ref}(t)-r(t)$,
the commanded inputs are
\begin{eqnarray}
\tau(t)&=&K_{p}^{(\theta)} e_\theta(t)+K_{i}^{(\theta)}\!\int_{0}^{t} e_\theta(s)\,ds+K_{d}^{(\theta)} \dot e_\theta(t),\\
f(t)&=&K_{p}^{(r)} e_r(t)+K_{i}^{(r)}\!\int_{0}^{t} e_r(s)\,ds+K_{d}^{(r)} \dot e_r(t).
\end{eqnarray}
Here the gains $K_{p,i,d}^{(\theta)}$ and $K_{p,i,d}^{(r)}$ are tuned for the angular and radial channels, respectively.

Under the actuator bounds in Eq.~(\ref{eq17}), we repeat the same initialization-error and input-disturbance tests with PID tracking. Figs.~\ref{figinitialerror} and \ref{figcontrolerror} show that feedback substantially improves terminal accuracy relative to the two open-loop protocols. The mean relative error ($\mathrm{MRE}$) is $1.344\%$ for initialization errors and $0.951\%$ for input disturbances; in our simulations the dominant contribution arises from residual deviations in the terminal velocities. This improvement comes at the cost of continuous sensing and real-time updates. Moreover, for a nonlinear coupled system, fixed gains may not be uniformly effective across operating conditions, and discrete-time implementation introduces a trade-off between tracking accuracy (small sampling time) and control activity/computational load.

\begin{figure}[t]
\centering
\includegraphics[width=\linewidth]{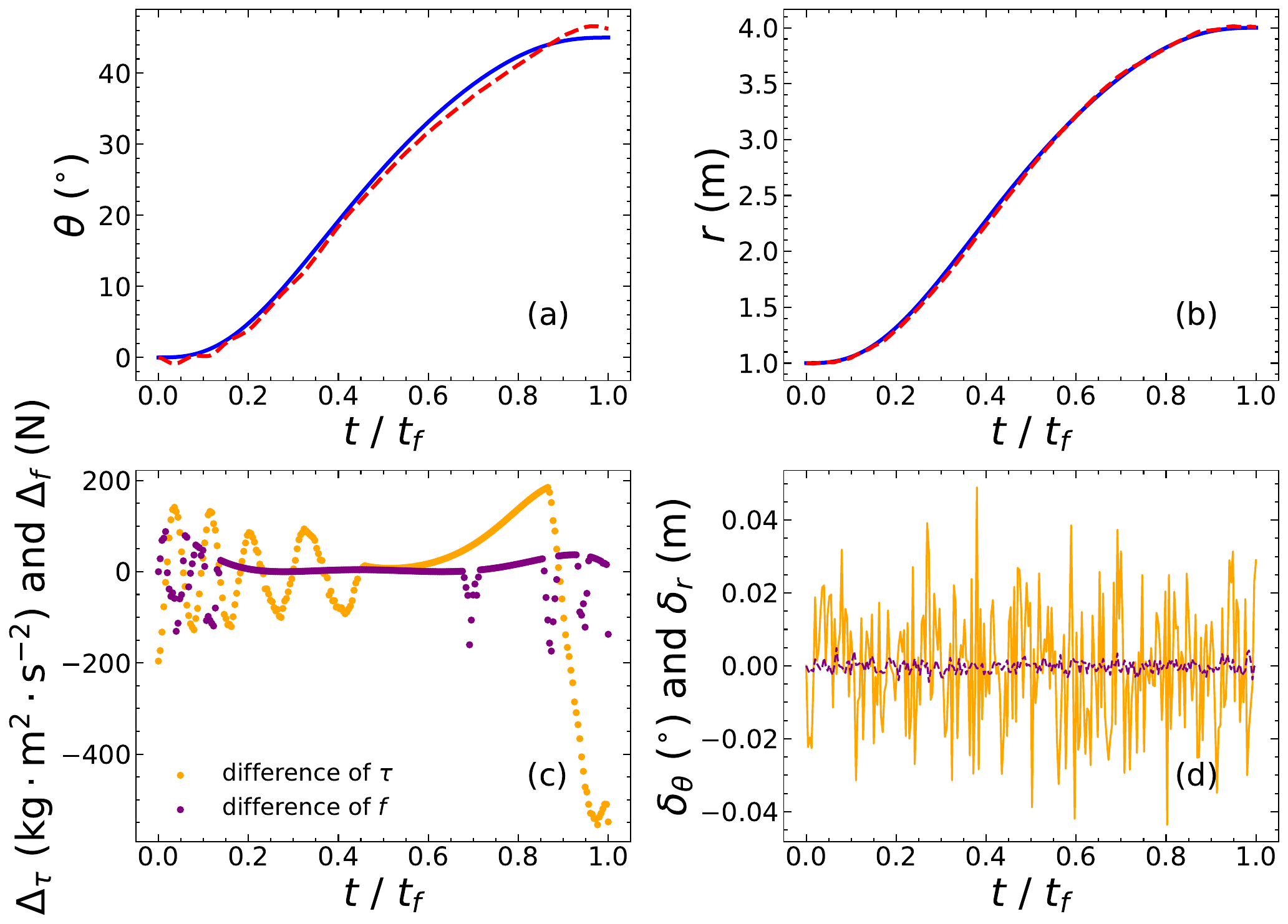}
\caption{\label{figmotionerror} The evolutions of $\theta(t)$ (a) and $r(t)$  (b)  for the reference trajectory (\ref{eqseventh}) (solid blue) and representative PID-tracked trajectories under measurement noise (dashed red).
(c) Differences $\Delta\tau(t)$ and $\Delta f(t)$ between PID-tracked inputs and the nominal open-loop inputs.
(d) Measurement noise added to $\theta$ (solid orange) and $r$ (dashed purple). The PID parameters and noise amplitudes are the same as in Fig.~\ref{figcontrolerror}.
}
\end{figure}

To assess sensitivity to monitoring errors, we add measurement noise at each sampling instant, $\delta_\theta(t)\in[-\pi/3600,\pi/3600]~\mathrm{rad}$ and $\delta_r(t)\in[-0.005,0.005]~\mathrm{m}$, so that the controller uses $[\theta(t)+\delta_\theta(t),\,r(t)+\delta_r(t)]$. Fig.~\ref{figmotionerror} shows a representative realization: the tracked trajectories remain close to the reference but become less smooth and develop visible oscillations. The corresponding feedback corrections, $\Delta\tau$ and $\Delta f$ in Fig.~\ref{figmotionerror}(c), exhibit pronounced discrete jumps compared with the smooth nominal STA inputs, reflecting rapid compensation based on noisy measurements. Over repeated trials with measurement noise, PID tracking yields an $\mathrm{MRE}$ of $1.559\%$, demonstrating the robustness gain of feedback at the price of increased control activity and reduced smoothness.

\subsubsection{Single-shot corrected STA}

Continuous feedback can strongly improve robustness but requires high-rate sensing and real-time computation and typically produces nonsmooth, high-frequency actuation. Motivated by the strong error suppression of time-optimal saturated inputs and by the robustness of PID tracking, we introduce an intermediate strategy: an STA protocol that remains open loop for most of the evolution but uses a single mid-course measurement to implement a short corrective action.

Specifically, we follow the nominal STA input designed from the seventh-order reference trajectory (\ref{eqseventh}) and take one measurement of the actual configuration $q(t_i)=[\theta(t_i),r(t_i)]^T$ at a chosen instant $t_i$. The measurement time is selected such that the initial mismatch is detectable while strong nonlinear amplification has not yet occurred. Based on this one-shot estimate, we apply a two-stage compensation that (i) reduces the accumulated position error and (ii) removes the velocity mismatch introduced during the correction, returning the system close to the nominal STA manifold before the final approach.

We define a normalized position error $\epsilon(t)=e(t)/(q_f-q_0)$, where $e(t)=[e_{\theta}(t),e_{r}(t)]^T$. The compensation durations are chosen as $t_1=|e(t_i)/v(t_i)|$ for position correction and $t_2=t_1/c_1$ for velocity compensation, where $v(t_i)$ is the reference velocity at $t_i$ and $c_1$ is a tunable parameter. The corrective input is implemented as
\begin{eqnarray}
Q(t)=\left\{\begin{matrix}
Q(t_i)\!\left[1+\dfrac{c_2}{t_1+c_3}\,\epsilon(t_i)\right], & t_i\le t<t_i+t_1,\\[6pt]
Q(t_i)\!\left[1-\dfrac{c_4}{t_2+c_5}\,\epsilon(t_i)\right], & t_i+t_1\le t<t_i+t_1+t_2,
\end{matrix}\right.
\end{eqnarray}
where $c_2$-$c_5$ are design parameters; $c_3$ and $c_5$ regularize the denominators, while $c_2$ and $c_4$ set the correction strengths in the two stages.
\begin{figure}[t]
\centering
\includegraphics[width=\linewidth]{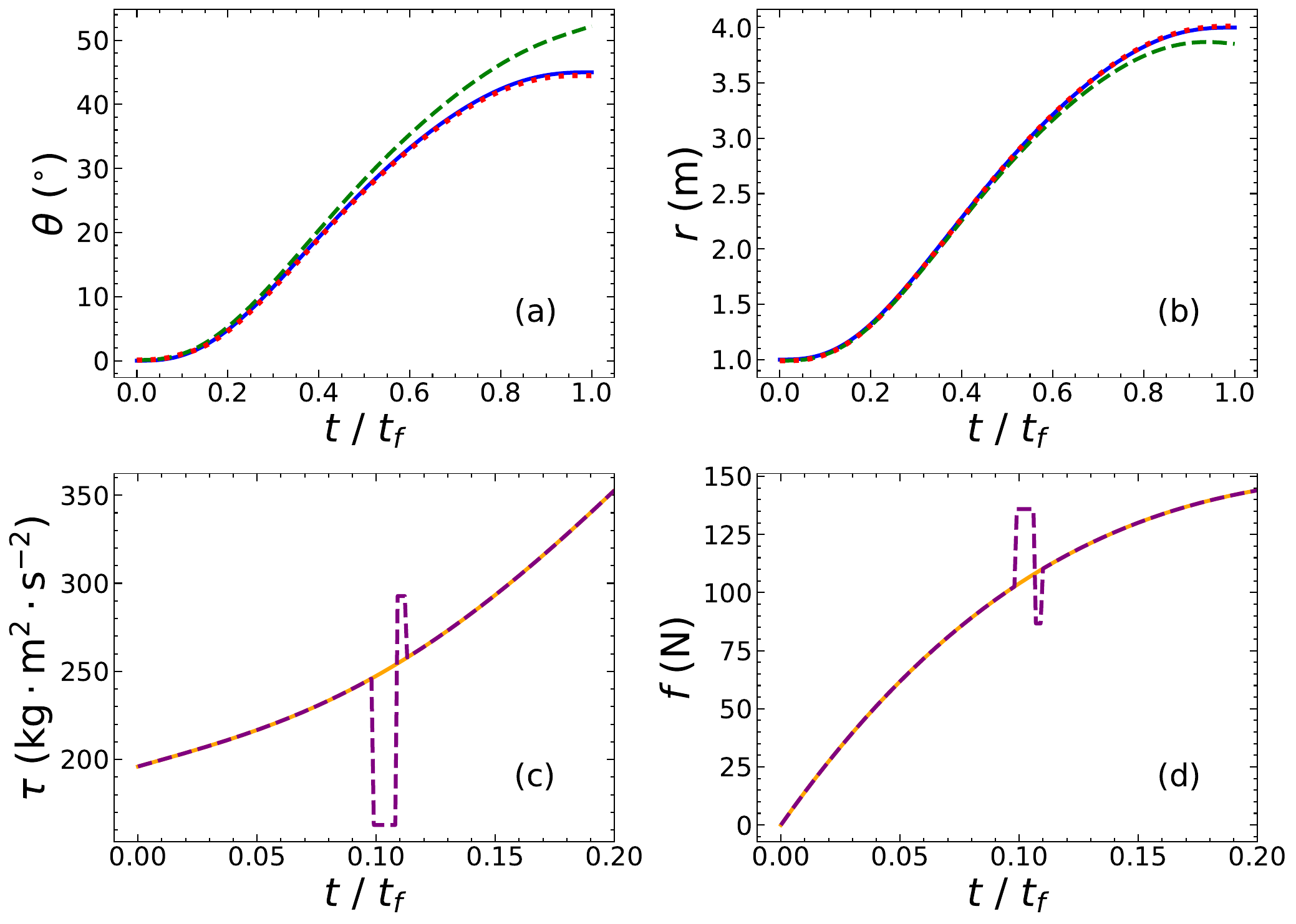}
\caption{\label{figoptpoly}(a) $\theta(t)$ and (b) $r(t)$ for the reference trajectory (\ref{eqseventh}) (solid blue), the nominal STA evolution (dashed green, $t_f=2.535~\mathrm{s}$), and the single-shot corrected STA evolution (dotted red).
(c) $\tau(t)$ and (d) $f(t)$ for the nominal (solid orange) and corrected (dashed purple) STA inputs.
Initial errors: $\delta_\theta(0)=\pi/1800~\mathrm{rad}$ and $\delta_r(0)=-0.01~\mathrm{m}$.
}
\end{figure}
We apply the corrected STA protocol under the same actuator constraints (\ref{eq17}). For the representative deviation $\delta_\theta(0)=\pi/1800~\mathrm{rad}$ and $\delta_r(0)=-0.01~\mathrm{m}$ shown in Fig.~\ref{figoptpoly}, the correction reduces $\mathrm{RE}$ from $8.401\%$ to $0.581\%$. Over the full grid of initial errors in Fig.~\ref{figinitialerror}, the resulting $\mathrm{MRE}$ is $0.891\%$, outperforming both the PMP time-optimal protocol and PID tracking in this setting.

Overall, the single-shot corrected STA provides a compromise between purely open-loop STA and fully closed-loop PID: it preserves the smooth nominal STA inputs for most of the evolution and avoids continuous feedback, yet achieves substantial error reduction via one measurement and a short segmented correction. Its limitations are that the correction parameters require tuning and the method remains sensitive to measurement noise at the monitoring instant.

{\section{Influence of dissipation}
\label{sec:dissipation}
Dissipation plays a nontrivial role in coupled nonlinear dynamics because it controls how transient kinetic energy generated by the $r$-$\theta$ coupling is removed during the motion. Although the previous sections focused on a representative choice of damping parameters, it is useful to examine to what extent the STA performance depends on the dissipation strength. For simplicity, we consider the quintic-polynomial STA protocol [Eq.~(\ref{eqquintic})] with $t_f=4~\mathrm{s}$ and scan the damping coefficients over the ranges
$B_1 \in [0,200]~\mathrm{kg\,m^2\,s^{-1}\,rad^{-1}}$, $B_2 \in [0,200]~\mathrm{kg\,s^{-1}}$.
For each pair $(B_1,B_2)$, we propagate the dynamics and evaluate the relative final-energy error $\mathrm{RE}$ defined in Sec.~\ref{STA&Optimization}.}

\begin{figure}[t]
\centering
\includegraphics[width=0.87\linewidth]{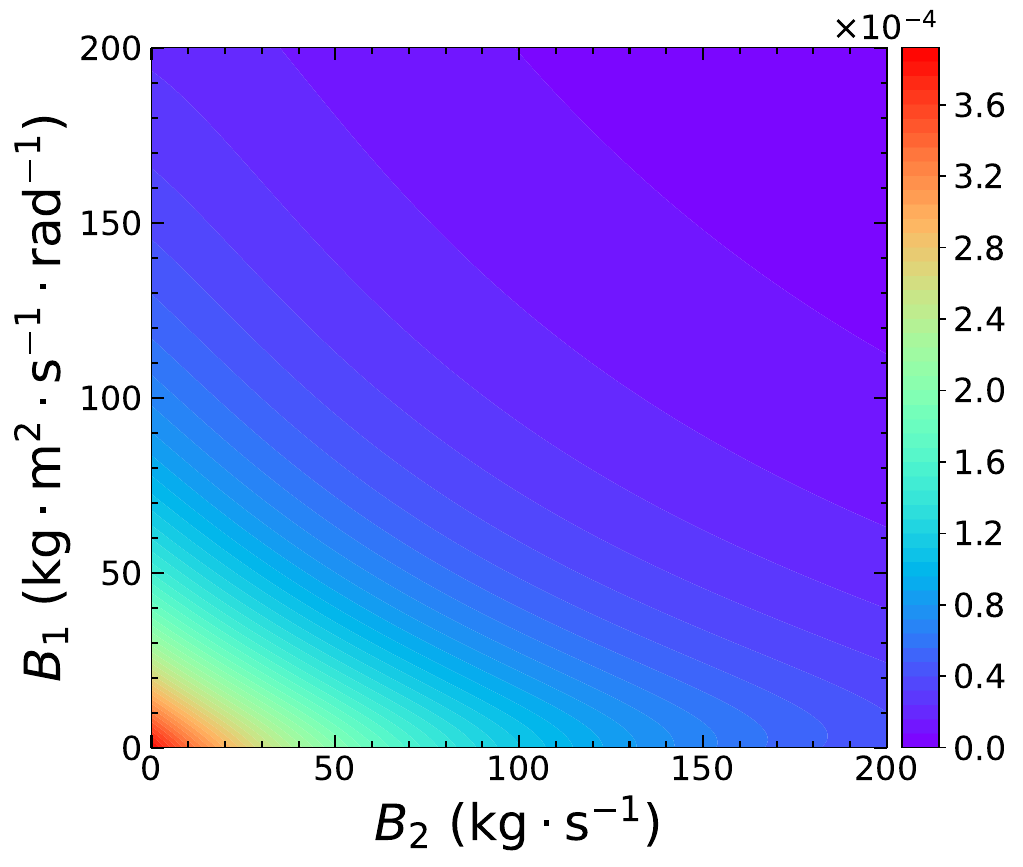}
\caption{\label{dissipationinfluenceRE}
Relative error $\mathrm{RE}$ of the final mechanical energy as a function of the damping coefficients $B_1$ and $B_2$ for the STA protocol based on the quintic trajectory (\ref{eqquintic}) with $t_f=4~\mathrm{s}$.
}
\end{figure}

{In the presence of viscous dissipation [Eq.~(\ref{eqdissipation})], the mechanical energy 
$E(t)=T(t)+V(t)$ satisfies
\begin{equation}
\dot E=-B_1\dot\theta^2-B_2\dot r^2+\tau\dot\theta+f\dot r,
\end{equation}
so that the dissipation terms $B_1\dot\theta^2$ and $B_2\dot r^2$ act as energy-removal channels for the transient kinetic energy generated during the motion. The results are shown in Fig.~\ref{dissipationinfluenceRE}.
The relative error decreases as the dissipation strength increases, indicating that stronger damping more efficiently removes the excess kinetic energy produced by the nonlinear $r$-$\theta$ coupling during the finite-time transfer. In this sense, dissipation acts as a relaxation channel that suppresses oscillatory remnants in the dynamics.
The variation is significant across the scanned parameter range. At the same time, even in the weakest-damping limit shown here, where the final-energy error is maximal, the relative error remains below $4\times10^{-4}$. This suggests that the quintic STA protocol remains robust over a broad range of dissipation strengths and that the main conclusions obtained from the STA protocols do not rely on a fine tuning of the damping parameters. 
Overall, although dissipation quantitatively affects the terminal accuracy, the qualitative behavior of the STA protocol remains robust against moderate variations of the dissipation parameters.}

\section{Conclusion and outlook}

Motivated by current efforts to transfer STA concepts from quantum dynamics to classical settings, we investigated a minimal yet representative coupled nonlinear Lagrangian system in polar coordinates with viscous dissipation. By prescribing boundary-conditioned trajectories for the generalized coordinates $q(t)$ and reconstructing the corresponding generalized torque and force through the EL equations, we obtained explicit open-loop driving profiles that realize fast point-to-point transfers. The endpoint conditions imposed on position, velocity, and acceleration suppress terminal excitations and provide a simple inverse-engineering route to smooth finite-time control in nonlinear dissipative dynamics.

Under actuator bounds, we compared three complementary strategies: (i) smooth inverse-engineered STA protocols, (ii) actuator-bounded time-optimal protocols obtained from Pontryagin’s minimum principle (PMP), and (iii) closed-loop PID tracking around the STA reference. Each approach highlights a distinct physical trade-off. Time-optimal solutions minimize the transfer time $t_f$ by exploiting near-saturated inputs and rapid switching, typically at the expense of nonsmooth actuation. PID feedback improves disturbance rejection by continuously correcting deviations, but requires high-rate sensing and real-time updates and may introduce increased control activity under noisy measurements. In contrast, inverse engineering provides a transparent route to smooth driving profiles, although purely open-loop protocols remain sensitive to mismatch and disturbances due to error propagation in the coupled nonlinear dynamics. The present comparison therefore clarifies how inverse engineering, actuator saturation, and feedback stabilization address different control objectives in nonlinear dissipative Lagrangian systems.

To bridge purely open-loop STA and continuous feedback, we introduced a single-shot corrected STA strategy based on one mid-course measurement followed by a short two-stage compensation. This correction preserves the smooth nominal STA profile for most of the evolution while mitigating the amplification of early deviations before the final approach. It therefore provides a practical compromise between robustness and implementation complexity, recovering part of the error suppression typical of strong-input or feedback-based schemes without continuous monitoring.

Finally, the present $r$-$\theta$ manipulator model also provides a simple classical analogue of inverse-engineering problems in quantum control. The variables $(r,\theta)$ coincide with the polar coordinates of a quantum particle in two dimensions (Appendix~\ref{app:quantum_analogue}), and the endpoint conditions used here mirror those employed in quantum STA to suppress nonadiabatic excitations. Beyond this illustrative model, the same inverse-engineering viewpoint can be extended to broader classes of nonlinear dissipative Lagrangian systems, including coupled pendula~\cite{Niu}, parametrically driven oscillators, and impact-like dynamics. Enlarging the admissible trajectory space may also provide a systematic route to incorporate additional design constraints, such as jerk limits, control-energy penalties, and bandwidth constraints, within a unified finite-time framework.

\begin{acknowledgments}
This work is supported by the National Natural Science Foundation of China (NSFC)  (12075145) and Science and Technology Commission of Shanghai Municipality (STCSM) (2019SHZDZX01-ZX04), PID2024-157842OA-I00 and PID2021-126273NB-I00 funded by MCIN/AEI/10.13039/501100011033 and by ``ERDF A way of making Europe'' and ``ERDF Invest in your Future'', the Basque Government through Grant No. IT1470-22, and the Severo Ochoa Centres of Excellence program through Grant CEX2024-001445-S. Y.B. acknowledges the Ayudas para contratos Ramón y Cajal (RYC2023-042699-I) and Projects in the field of Artificial Intelligence 2025 (AIA2025-163435-C44). 
\end{acknowledgments}

\appendix

\section{Quantum analogue of the $r$-$\theta$ manipulator}
\label{app:quantum_analogue}

It is convenient to interpret the $r$–$\theta$ dynamics as the classical (Ehrenfest) limit of a driven quantum particle moving in two dimensions and described in polar coordinates \cite{AnderQST,Simsek,XiaojingQuantum}. A minimal controllable setting is a particle of mass $m$ confined by a separable, time-dependent trap composed of a radially translated harmonic well and an angular potential whose minimum can be rotated in time,
\begin{equation}
V(r,\theta,t)=\frac{1}{2}m\omega_r^2(t)\big[r-r_0(t)\big]^2+U_\theta\big[\theta-\phi(t),t\big].
\end{equation}
Here $r_0(t)$ is a radial setpoint, $\omega_r(t)$ controls the radial confinement, and $\phi(t)$ provides an angular setpoint through a time-dependent rotation of the angular potential.

The corresponding Hamiltonian is
\begin{equation}
H(t)=\frac{p_r^2}{2m}+\frac{L_z^2}{2mr^2}+V(r,\theta,t),
\qquad L_z\equiv p_\theta,
\label{eq:H_polar}
\end{equation}
where $p_r$ is the radial momentum and $L_z$ is the canonical angular momentum. In a semiclassical (Ehrenfest) sense, the evolution of coordinate expectations is governed by the classical Hamilton equations:
$\dot r=\partial H/\partial p_r$, $\dot\theta=\partial H/\partial L_z$,
$\dot p_r=-\partial H/\partial r$, and $\dot L_z=-\partial H/\partial\theta$.
The nonlinear coupling between $r$ and $\theta$ is purely geometric and originates from the kinetic term in polar coordinates.

Using $\dot r=\partial H/\partial p_r=p_r/m$ and differentiating once more, we obtain
\begin{equation}
m\ddot r=\dot p_r=\frac{L_z^2}{mr^3}-\partial_r V(r,\theta,t).
\end{equation}
Since $L_z=mr^2\dot\theta$, the centrifugal contribution becomes $L_z^2/(mr^3)=mr\dot\theta^2$, leading to
\begin{equation}
m\ddot r-mr\dot\theta^2=-\partial_r V(r,\theta,t).
\label{eq:radial_fromV}
\end{equation}
Similarly, differentiating $L_z=mr^2\dot\theta$ yields
\begin{equation}
\dot L_z=m\big(2r\dot r\dot\theta+r^2\ddot\theta\big),
\end{equation}
and using $\dot L_z=-\partial_\theta V$ one obtains
\begin{equation}
mr^2\ddot\theta+2mr\dot r\dot\theta=-\partial_\theta V(r,\theta,t).
\label{eq:theta_fromV}
\end{equation}
Eqs.~(\ref{eq:radial_fromV}) and (\ref{eq:theta_fromV}) reproduce the characteristic coupling terms in the classical model, namely $-mr\dot\theta^2$ and $2mr\dot r\dot\theta$.

The classical force and torque channels correspond to potential gradients,
\begin{eqnarray}
f(t)\sim -\partial_r V(r,\theta,t),\qquad
\tau(t)\sim -\partial_\theta V(r,\theta,t),
\label{eq:force_torque_map}
\end{eqnarray}
and for the radial trap term $\tfrac{1}{2}m\omega_r^2(t)\big[r-r_0(t)\big]^2$ one obtains explicitly
\begin{eqnarray}
f(t)&=&-m\omega_r^2(t)\big(r-r_0(t)\big),\\
\tau(t)&=&-\partial_\theta U_\theta\big[\theta-\phi(t),t\big].
\label{eq:force_torque_explicit}
\end{eqnarray}

To reproduce the gravitational terms used in the main text,  we add
$V_g(r,\theta)=mgr\sin\theta$,
so that $\partial_\theta V_g=mg r\cos\theta$ and $\partial_r V_g=mg\sin\theta$. With the total potential $V_{\rm tot}=V+V_g$ and with additional viscous damping terms (nonconservative and thus not derivable from a potential), the resulting classical equations take the same form as Eqs.~(\ref{eq:theta_dyn}) and (\ref{eq:r_dyn}). This makes explicit that the strong nonlinear coupling in the classical $r$-$\theta$ model is
the direct classical counterpart of the kinetic coupling inherent to the polar-coordinate Hamiltonian~(\ref{eq:H_polar}). In this sense, prescribing boundary-conditioned trajectories for $(r(t),\theta(t))$ and inverting the equations of motion parallels quantum STA inverse engineering, where time-dependent Hamiltonians are designed to satisfy terminal conditions without unwanted excitations.

If the angular potential minimum is rotated by $\phi(t)$, it is often convenient to transform into the co-rotating frame with
\begin{equation}
U(t)=\exp\!\left[-\frac{i}{\hbar}\phi(t)L_z\right].
\end{equation}
The rotating-frame Hamiltonian
\[
H_{\rm rot}(t)=UH(t)U^\dagger-i\hbar\,U\dot U^\dagger
\]
becomes
\begin{equation}
H_{\rm rot}(t)=\frac{p_r^2}{2m}+\frac{L_z^2}{2mr^2}
+\frac{1}{2}m\omega_r^2(t)\big[r-r_0(t)\big]^2
+U_\theta(\theta,t)
-\Omega(t)L_z,
\end{equation}
with $\Omega(t)\equiv\dot\phi(t)$. The Coriolis-like term $-\Omega(t)L_z$ acts as an explicit, controllable angular drive. For small angular excursions, one may further approximate
\begin{equation}
U_\theta(\theta,t)\simeq \frac{1}{2}I(t)\omega_\theta^2(t)\theta^2,
\qquad I(t)=mr^2,
\end{equation}
highlighting the role of $I(t)$ as an effective moment of inertia, consistent with the classical $r$-$\theta$ description.

\end{document}